\documentclass[useAMS,usenatbib,usegraphicx]{mn2e}

\begin{document}

\title[Dynamical phasing of Type II Cepheids]{Dynamical phasing of
Type II Cepheids}
\author[J.\ A.\ McSaveney et al.]{J. A.
McSaveney,$^{1}$$^{2}$$^{3}$\thanks{Email (respectively):
jennifer.mcsaveney@mso.anu.edu.au,
karen.pollard@canterbury.ac.nz,
peter.cottrell@canterbury.ac.nz} K.\ R.\ Pollard,$^{1}$
P.\ L.\ Cottrell$^{1}$\\
$^{1}$Mount John University Observatory, Department of Physics and
Astronomy, University of Canterbury, Christchurch, New Zealand \\
$^{2}$Research School of Astronomy and
Astrophysics, Australian National University, Canberra, Australia \\
$^{3}$Center for Stellar and Planetary Astrophysics, Department of
Mathematical Sciences, Monash University, Melbourne, Australia \\
}

\date{Accepted 2005?.
        Received 2005 ?;
        in original form 2004 ?}

\pagerange{\pageref{firstpage}--\pageref{lastpage}} \pubyear{2004}

\maketitle

\label{firstpage}

\begin{abstract}

In this paper we examine the problems of phasing using light curves
and offer an alternate technique using the changes in acceleration to
establish the zero point. We give astrophysical justification as to
why this technique is useful and apply the technique to a selection of
Type~II Cepheids. We then examine some limitations of the technique
which qualify its use.

\end{abstract}

\begin{keywords}
stars: oscillations --- techniques: miscellaneous.
\end{keywords}

\section{Introduction}

When studying intrinsically pulsating stars
a key challenge is to find common physical processes at a given
pulsational phase in stars of different spectral characteristics and
periods. In order to do this, a common analysis technique is required
which can find links between apparently disparate groups of
objects. This is especially true when trying to understand what the
stars are physically doing by linking spectroscopic
features to the photometry. Uncertainties exist when using the
photometry to establish the zero point, as it cannot be guaranteed
that the same spectroscopic features are occurring at the same
photometric phase between different stars. This leads to ambiguities
in comparisons between stars. An alternative approach is required
that allows a clearer comparison and a better understanding of
the physical processes involved.

In this paper we examine the problems of phasing using light curves. We
offer an alternative technique using the minimum of the acceleration
curve as the zero point. We give astrophysical
justification as to why this technique is useful and apply the
technique to a selection of stars. Some of the
limitations of the technique are then examined.

\begin{figure*}
\includegraphics[width=0.98\textwidth]{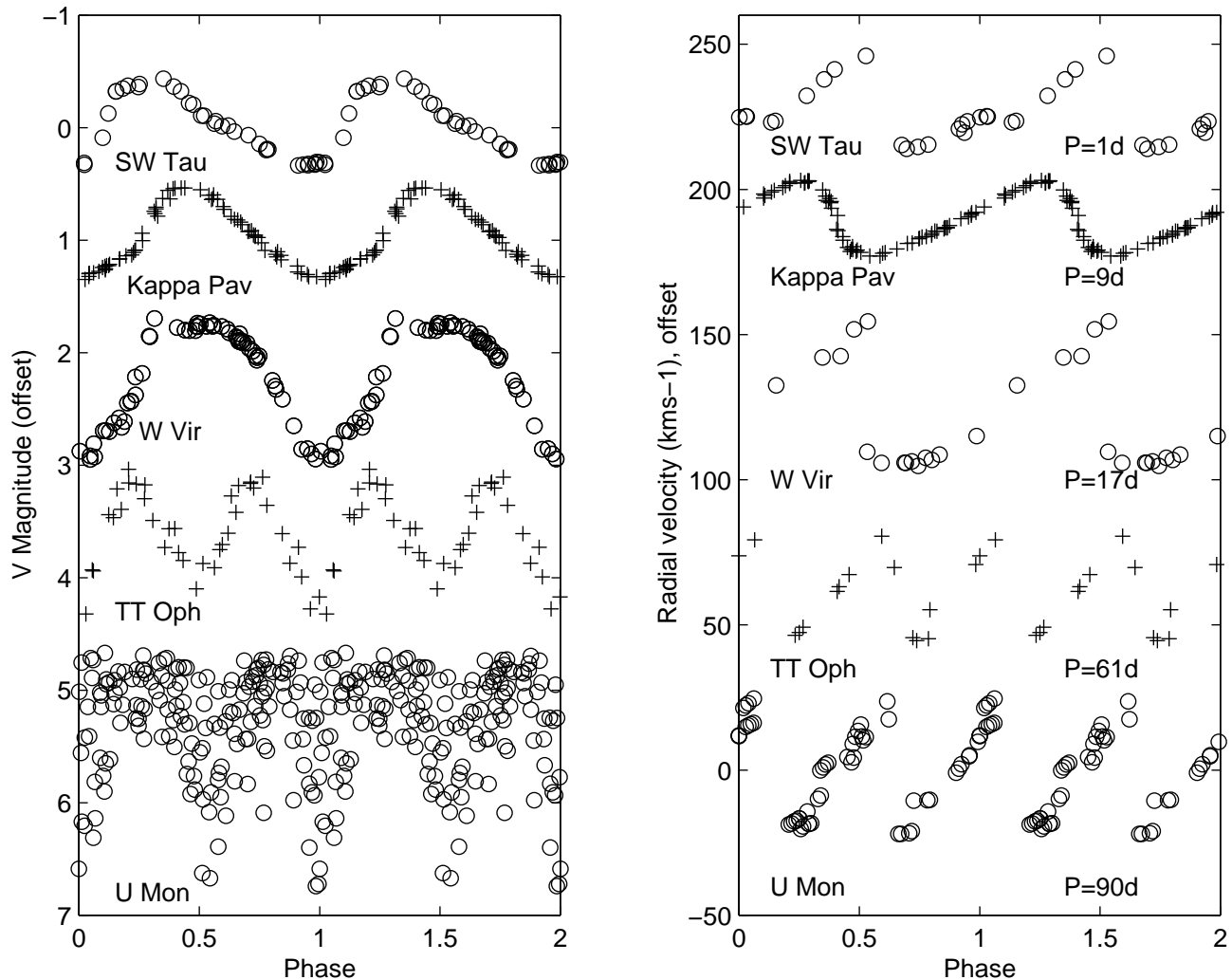}
\caption{Photometry (left) and velocity curves (right) of selected
Type~II Cepheids phased using a zero point of minimum $V$
magnitude.}
\label{figrvphotostack}
\end{figure*}

\section{Traditional phasing}

In order to be able to understand the regular pulsations of stars one
needs to characterize the basic parameters of what changes over a
pulsation cycle. One method is to plot the data (such as photometry
  or radial velocity) as a function of pulsational phase. This is shown
in Figure \ref{figrvphotostack} for a selection of Type II Cepheids, 
with periods
ranging from 1.6 d (SW~Tau) to 90 d (U~Mon). In this, using a fixed
zero point and a known period, data over a wide time interval can be
compared. Traditionally, the zero point of this phasing has been based
on the time of maximum
\citep[e.g.][]{moffett1984,schmidt2003,schmidt2003ii}, or minimum
\citep{pollard1996} of the star's light curve. This is due to the
historical precedent that stellar variability is generally first
discovered through overall light variations. In the more regular
Cepheids such as SW~Tau and $\kappa$~Pav (top, Figure
\ref{figrvphotostack}), the determination of the position of an
extremum is a relatively straight-forward process. However the more
complicated RV~Tauri stars (i.e. U~Mon, lowest curve, Figure
\ref{figrvphotostack}), are less obvious. There is an additional
problem that the phases of maximum and minimum light do not
necessarily occur at the same time in different photometric filters. As such,
while the extrema in the light curves represent a relatively easily
observed quantity, they do not represent an optimal means by which to
compare the astrophysical behaviour of the stars.
Two main problems are encountered in phasing data based on maximum
or minimum light. Firstly, a photometric filter region must be
chosen, and secondly, the light extrema must be well defined.

\begin{figure}
\includegraphics[width=0.48\textwidth]{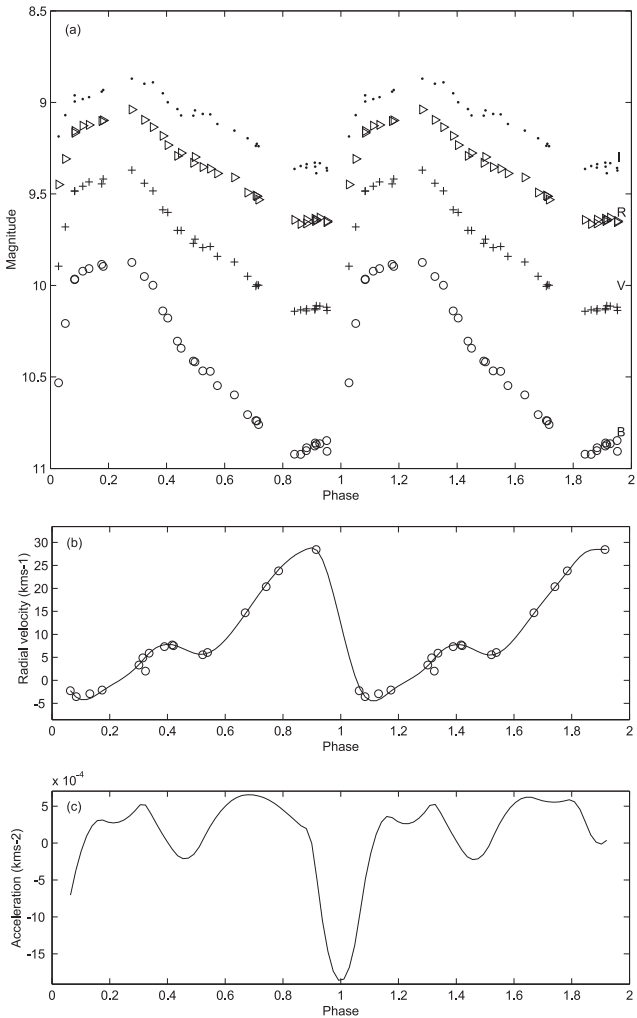}
\caption{SW~Tau (a) {\it BVRI}
    photometry, (b) spline-fitted mean Fe~I radial velocities, and (c)
    acceleration curves. SW~Tau is phased on a 1.58356-day
    period from \citet{mcsaveney2003} and using the zero point from the
    acceleration curve.}
\label{figswtau}
\end{figure}

\begin{figure}
\includegraphics[width=0.48\textwidth]{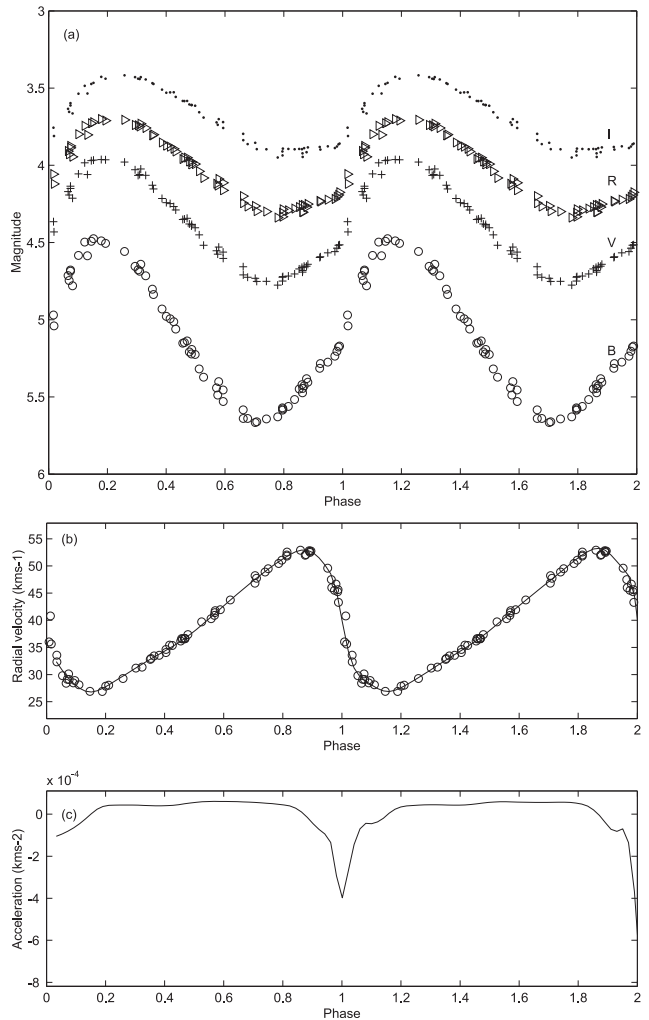}
\caption{$\kappa$~Pav (a) {\it BVRI}
    photometry, (b) spline-fitted mean Fe~I radial velocities, and (c)
    acceleration curves. $\kappa$~Pav is phased on a 9.0714-day period
    from \citet{mcsaveney2003} and using the zero point from the acceleration
    curve.}
\label{figkappapav}
\end{figure}

\begin{figure}
\includegraphics[width=0.48\textwidth]{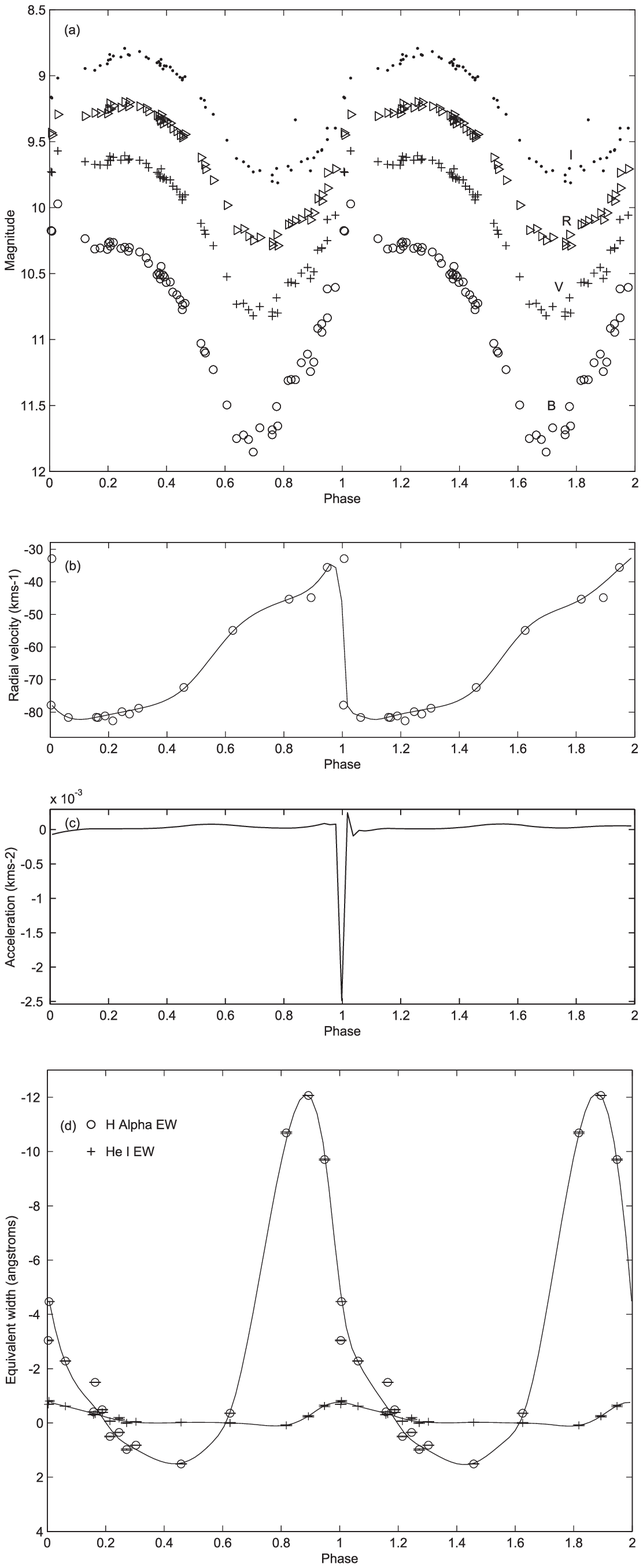}

\caption{W~Vir (a) {\it BVRI}
    photometry, (b) spline-fitted mean Fe~I radial velocities, (c)
    acceleration curves, and (d) H$\alpha$ and He~I equivalent widths. W~Vir
    is phased on a 17.2768-day period from \citet{mcsaveney2003} and using
    the zero point from the acceleration curve.}
\label{figwvir}
\end{figure}

The first is trivial only if all the photometry used is from a single
band-pass. As soon as other band-passes are observed, the extrema do
not exactly coincide because they are sampling different regions of
the stellar atmosphere. This is clearly seen in Figures
\ref{figswtau}a, \ref{figkappapav}a and \ref{figwvir}a and Table
\ref{tabphases}, where maximum light shifts to a later phase at longer
wavelengths (from {\em B} to {\em I}). Astrophysically, this is due to
changes in temperature and hence flux distribution over the pulsation
cycle. As the stellar photosphere expands and cools, the peak flux
shifts to longer wavelengths, causing maximum light in the redder
band-passes to occur slightly later in the pulsation cycle. An extreme
case is W~Vir (Figure \ref{figwvir}a and Table \ref{tabphases}) in
which maximum light shifts forward in phase by 0.2 of a pulsation
cycle between {\em V} and {\em R} as the primary maximum becomes less
prominent in the red. This maximum is most likely associated with the
shock wave propagating through the star during this phase
\citep{abt1960}.

The second problem, an ill-defined maximum or minimum, will depend
critically on the data density
obtained and the repeatability of the light curve. As with any fitting
to observational data, a large scatter in the data points can introduce
large errors
in the extrema fitting. The broadness of the extrema can also add to
the difficulty in determining the zero point. This is seen
in SW Tau (Figure \ref{figswtau}) where the minima are quite broad
in all the band-passes observed. As seen from Figure
\ref{figrvphotostack}, this represents one of the more regular stars
of the sample. The scatter for U~Mon is even greater and is affected
by long term variations in the light curve and
irregularities in the depths of the minima.

Both of these problems lead to slight shifts in the zero points
between the stars, such that the same phase does not represent the
same physical state for each star. It should also be noted that
different stars spend a different fraction of their pulsation
cycle brightening from minimum to maximum light. Hence, when basing a
zero point on minimum light, the phases of maximum light compared are
offset and vice versa.This can been seen in
Figures \ref{figswtau}a, \ref{figkappapav}a and \ref{figwvir}a and
Table \ref{tabphases}, where
the duration of the rise from minimum to maximum light ranges from 0.2 to
0.5 of a pulsation cycle. This leads to difficulty in comparing stars
with a range of light curves at the same phases. 

\begin{table*}
\begin{minipage}{170mm}
\caption{Phase shifts$^{a}$ (columns 2 to 9) of photometric extrema
relative to minimum {\em V} light, which by definition is
phase=0.0. Columns 10 and 11 are respectively the phase shift between
photometric and dynamic zero points ($\Delta$) and its uncertainty
(HWHM). See text (end of Section \ref{sec:altphase}) for detailed
explanation of these terms.} 
\begin{center}
\label{tabphases}
\begin{tabular}{cccrrcccccc}
\hline
Star & $\phi_{min}$ \em{B} & $\phi_{min}$\em{V} & $\phi_{min}$ \em{R}
& $\phi_{min}$ \em{I} & $\phi_{max}$ \em{B} & $\phi_{max}$ \em{V} &
$\phi_{max}$ \em{R} & $\phi_{max}$ \em{I} & $\Delta
\phi_{V_{min}-dynamic}$
& HWHM\\
\noalign{\smallskip}
\hline
\noalign{\smallskip}
SW~Tau       & $-0.13$ & 0.00 &$-0.12$~ & $-0.01$~ & 0.21 & 0.35 &
0.36 & 0.37 & 0.07 & 0.09 \\
$\kappa$~Pav & $-0.02$ & 0.00 &$0.02$~  & $0.03$~  & 0.42 & 0.44 &
0.46 & 0.50 & 0.25 & 0.03 \\
W~Vir        & $-0.01$ & 0.00 &$0.01$~  & $0.02$~  & 0.34 & 0.36 &
0.56 & 0.56 & 0.29 & 0.01 \\
TT~Oph       & $-0.02$ & 0.00 &$0.01$~  & $0.02$~  & 0.20 & 0.72 &
0.23 & 0.24 & 0.12 & 0.10 \\
U~Mon        & $-0.01$ & 0.00 &$0.02$~  & $0.04$~  & 0.67 & 0.27 &
0.79 & 0.76 & 0.15 & 0.04\\
\hline
\noalign{\smallskip}
\end{tabular}
\end{center}
$^{a}$The uncertainty in these phase shifts
varies from star to star and between passbands. The range is from
$\pm$0.07 (for {\em V} in W~Vir) to $\pm$0.16 (for {\em I} in $\kappa$
Pav) when measured at 0.1 mag above the minimum (or below the maximum) of any
light curve.
\end{minipage}
\end{table*}

An alternative approach which phases the astronomical data based on
the same physical state for each star is desirable.

\section{Alternate phasing techniques}
\label{sec:altphase}

The pulsation cycle of a star can be observed physically in
a number of ways through changes in brightness (photometry), radial velocity
(spectroscopy) and angular diameter
(interferometry). Having discussed the problems of using the light
curves, we turn to the other techniques. Both
give an indication of radii changes, though the angular diameter far more
directly than the radial velocities.
Maximum, mean or minimum radius would be useful
as a calibration point. To date Type II Cepheids have not been
observed interferometrically, so the
radial velocities remain the most easily observed parameter. Ideally,
one would use radial velocity curves to obtain radii changes and hence
well defined zero points for phasing the data. However, this is not a 
straight-forward process.

Derivation of radii changes from velocity curves requires integration
of the pulsational velocity curve. This differs from the measured
radial velocity curve due to limb-darkening effects across the disk and
the fact that the star is a spheroidal object changing in size and not
a point source moving towards and away from us. This can be corrected
for by the use of a geometric projection and limb darkening
coefficient {\it p}, but the exact value of the coefficient varies in the
literature (1.31, \citet{hindsley1986}, 1.41,
\citet{albrow1994b}). As well as varying from star to star, this
``constant'' also varies as a function of pulsational phase, method of
velocity measurement and line formation depth
\citep{sabbey1995,nardetto2004}. These values and statements are also
based on discussions of Classical Cepheids and their applicability to the
Type II Cepheids is not proven.
In addition, the radius derived from
a particular line or species will trace the radius of that particular line
formation region. Certain infra-red regions will probe deeper into
the stellar photosphere and produce different radii to those
determined from iron lines in the visual wavelength regions.

Minimum radius of the star has been used previously for phasing
\citep{marengo2002} by setting the zero phase point to be the time at
which the pulsational velocity moved from a positive to a negative
value. This does, however, require the transformation of the radial
velocity into a pulsational velocity, with the uncertainties mentioned
earlier. In a more recent paper, \citet{marengo2004} advocated 
dynamical phasing
based on the radial velocity curve. The reasoning
given was this phasing was based on hydrodynamical quantities
and could thus be more readily compared with the
time-dependent hydrodynamical models. These authors also discussed the
phase lag observed using both the
light maximum and radius minimum zero points.

Given the provisos of working with the pulsational velocities and
radii, it was decided to examine the acceleration curve
produced by taking the derivative of the radial velocity curve. While
not the true acceleration curve, as it is not the derivative of the
pulsational velocity, it is a clear indication of changes in the
bulk motion of the line formation region. The minimum of this
acceleration was then used, as for these stars it marked the
transition between the bulk of the material in the line formation
region falling inwards and moving outwards. By using the same species
from the same wavelength regions, a more robust comparison of material
under the same temperature and pressure conditions is obtained for the
different stars. This also minimized the number of assumptions
required in analyzing the data and is model independent.

In applying this technique a spline curve is fitted to the phased
radial velocity curve of a particular line or species (see Figures
\ref{figswtau}b, \ref{figkappapav}b and \ref{figwvir}b). Care is taken
to fit over several repeated cycles to avoid end effects. This curve
is then differentiated to find the acceleration. The minimum of this
curve defines the phasing zero point, namely minimum acceleration.
The data are then rephased based upon this. Figures \ref{figswtau}c,
\ref{figkappapav}c and \ref{figwvir}c show that the acceleration
minimum is far more clearly defined than the photometric minima shown
in Figures \ref{figswtau}a, \ref{figkappapav}a and \ref{figwvir}a.

Quantitative estimates of the uncertainty of the derived dynamical
zero point phase have been determined from the acceleration data,
based upon the half width at half the minimum of the acceleration dip
(see HWHM in Table 1). We note that it is not just dependent upon the
data density around the minimum acceleration point, but also depends
upon the steepness of the radial velocity curve at that point.

\section{Application}

To test this technique of dynamical phasing, it has been applied to a
selection of Type II Cepheids. The observations are taken from
McSaveney (2003). The {\em BVRI} photometry is from observations made by
A.C.\ Gilmore and P.M.\ Kilmartin, using automated, single-channel photometers
mounted on the Optical Craftsmen 0.61-m telescope at Mount John
University Observatory (MJUO). The velocities are measured from
spectra obtained by JAM, using a high-resolution spectrograph mounted
on the McLellan 1.0-m telescope at MJUO. Spectroscopic data were reduced
using FIGARO processes which were supplemented by MATLAB scripts.

The velocity curves shown here were obtained by averaging a selection of
Fe~I line velocities. These Fe~I lines, with a range of excitation
potentials and line strengths, represent a region of the star's
photosphere in which line formation is well understood and can be
considered representative of the star. We are aware of line level
effects between various species (see for example
\citet{wallerstein1992}), but in this paper we are considering only
those layers representing these Fe~I lines.

The individual velocities were measured by taking a mean of the line
bisector velocities at depths 0.7, 0.8 and 0.9 for each line. This
followed the line bisector method as developed by \citet{albrow1994}
and used in \citet{wallerstein1992}. This was used in preference to
methods using line core velocities, to smooth out any
line asymmetries. These line asymmetries were outside the scope of 
this work \citep{mcsaveney2003}.

The three stars shown in this paper (SW~Tau -- Figure \ref{figswtau},
$\kappa$~Pav -- Figure \ref{figkappapav} and W~Vir -- Figure
\ref{figwvir}) constitute a representative sample of Type II Cepheids
of periods ranging from 1.58 to 17.28 days. The Fe~I lines from which these
velocities were measured show a range of behaviour over a pulsational
cycle, from little change in line width in SW~Tau (the shortest
period star), to mild broadening at phase 0.0 in $\kappa$~Pav, to extensive
broadening and line-splitting at phase 0.0 in W~Vir.
As can be seen from Figures \ref{figswtau}, \ref{figkappapav}
and \ref{figwvir}, the
dynamical phasing technique can be used on all of these stars,
despite the differences
in velocity curve produced by the different line behaviours.

The acceleration curves shown for each star (Figures
\ref{figswtau}c, \ref{figkappapav}c and \ref{figwvir}c) all have
distinct minima that have been rephased to 0.0. Even in the case of
SW~Tau, where a second minimum is found around phase 0.4--0.5 due to a
bump in the velocity curve, this is easily distinguished from
the primary velocity field reversal.

The result of this phasing provides a much easier comparison between these
stars, as it is clear when the motions of the velocity fields are
reversing. In contrast, by phasing using {\em V} minimum (Figure
\ref{figrvphotostack}), it occurs between phase 0.85 and 0.95 in
SW~Tau, at $\sim$0.8 for $\kappa$~Pav and $\sim$0.7 in
W~Vir. These are shown quantitatively in Table 1, where the phase
shifts between V minimum and dynamical zero points ($\Delta$) are given
for the five Type~II Cepheids in our sample. These $\Delta$s show that
there is more than 0.2 in phase difference between these values that
is not related in any way to a physical stellar parameter. Also the
dynamical zero points have a much better defined position (HWHM in
Table 1 is between 0.01 and 0.10) when compared to the uncertainties in
the photometric zero points, which range from 0.07 to at least 0.16 and
vary considerably from passband to passband for a given star. This
effect is also clearly visible in Figures \ref{figswtau}, \ref{figkappapav}
and \ref{figwvir}, where the width
(in phase) of the acceleration minimum for a given object is in all
cases much narrower than any of its respective photometric passband
minima. 

Astrophysically, this shows that the dynamical zero point provides a
clear common phase compared to that indicated by any particular
photometric bandpass. The dynamical zero point indicates when the 
direction of motion of the line
formation region shifts from falling inwards to moving outwards. In
these stars, this is when shock-wave features are
detected as a shock propagates through the line forming regions. Hence
spectral features such as line broadening, splitting and emission all
occur around this phase. This is seen in the H$\alpha$ and He~I
equivalent widths (see Figure \ref{figwvir}d) where the strongest emission
occurs at around phase 0.0. This is also observed in a wider range of Type II
Cepheids (McSaveney, 2003).

\section{Discussion}

There are several assumptions which will have an affect on the results
obtained when using this technique. These include
what line species are used, the method of measuring the
radial velocities and how the spline curve is fitted.

An important point to note with this technique is that different line
species are formed over slightly different regions of the stellar
photosphere, and will therefore give slightly different zero
points. While this may be construed as a problem as large as that of
the shifts in zero point from the photometry, this is not actually the
case. The spectral line formation regions can be far more specifically
defined than the far broader photometric regions, particularly for the
dominant species (usually Fe~I), from which the velocities are determined.
When establishing the velocities care should be taken to use lines of
similar species,
excitation potential and log({\it gf}) such that there is a high
degree of confidence that the lines are formed in the same region of
the stellar photosphere. Averaging H$\alpha$ velocities with Fe~I
velocities would be counterproductive given the large phase lags and
amplitude differences observed between these lines in Type~I and
Type~II Cepheids
\citep{wallerstein1992,vinko1998,petterson2002,mcsaveney2003}.

The radial velocity measuring technique will also affect the velocity
determined. Techniques fitting to the sides of the lines, or taking the
line bisector value from midway up or higher (closer to the star's
continuum) in the line, will be more
easily distorted by line broadening or splitting. This will result in
different amplitudes. Techniques fitting
to the line bisector from deeper in the line core (as used in this
work) give a better indication of the maximum and minimum velocities
achieved by material in the line forming regions. The crucial point 
is to be consistent in
the manner of application such that it is clear that the stars (as
shown by the appropriate spectral lines) show the same measured
zero-point behaviour.

As with any data set, good radial velocity phase coverage is essential,
particularly at the critical phases between minimum and maximum light
when the minimum radius occurs. To avoid distortion by end effects in the
spline fitting, repetition of the full velocity curve is required. This
technique is quite data intensive. However, for the comparison of
stars with a range of periods all with large datasets, it is invaluable.

\section{Conclusion}

Using the acceleration changes as a zero point in dynamical phasing offers a
substantial improvement over previous phasing techniques in understanding the
pulsations in Type~II Cepheids. It shifts the phasing emphasis from the
more ambiguous light extrema to the astrophysically more specific, yet
still observable, changes observed through radial velocity
curves. This allows a better understanding of the physics of the
stellar pulsations and clearer comparison between stars, as they can
be collectively phased to a more clearly defined astrophysical state.

This technique has been applied to a selection of Type~II Cepheids and
found to work extremely well for a variety of velocity curve shapes and
periods. From a preliminary
survey of SW~Tau, $\kappa$~Pav and W~Vir, it is clear that the
shock-associated features of line broadening and line splitting, and
H$\alpha$ and He~I emission, are associated with the bulk motion
reversal observed and can be more clearly compared by using this
particular zero point. More detailed study of the shock behaviour
for a wider range of Type~II Cepheids using this technique can be found in
McSaveney et al. (2005, in preparation).

\section*{Acknowledgments}

This work is based on part of JAM's PhD thesis at the University of
Canterbury. JAM was financially supported in her PhD thesis through a
Department of Physics and Astronomy Teaching Scholarship (1999--2001),
the B.G.\ Wybourne Scholarship (1999), the Dennis William Moore
Scholarship (2001) and an Amelia Earhart Fellowship (2001) from the
Zonta Foundation.

This research has used the SIMBAD data base operated at CDS,
Strasbourg, France, the Vienna Atomic Line Database and the NASA
Astrophysics Data System database.

\bsp

\label{lastpage}


\begin{thebibliography}{99}
\bibitem[\protect\citeauthoryear{Abt \& Hardie}{1960}]{abt1960} Abt
H.A., Hardie R.H., 1960, ApJ, 131 155
\bibitem[\protect\citeauthoryear{Albrow}{1994}]{albrow1994} Albrow
M.D., 1994, PhD thesis, University of Canterbury
\bibitem[\protect\citeauthoryear{Albrow \&
Cottrell}{1994}]{albrow1994b} Albrow M.D., Cottrell P.L., 1994, MNRAS,
267, 548
\bibitem[\protect\citeauthoryear{Hindsley \&
Bell}{1986}]{hindsley1986} Hindsley R., Bell R.A., 1986, PASP, 98, 881
\bibitem[\protect\citeauthoryear{Marengo et al.}{2002}]{marengo2002}
Marengo M., Sasselov D.D., Karovska M., Papaliolios C., Armstrong
J.T., 2002, ApJ, 567, 1131
\bibitem[\protect\citeauthoryear{Marengo et al.}{2004}]{marengo2004}
Marengo M., Karovska M., Sasselov D.D., Sanchez M., 2004, ApJ, 603, 285
\bibitem[\protect\citeauthoryear{McSaveney}{2003}]{mcsaveney2003}
McSaveney J.A., 2003, PhD thesis, Univ.\ of Canterbury
\bibitem[\protect\citeauthoryear{Moffett \&
Barnes}{1984}]{moffett1984} Moffett T.J., Barnes T.G., ApJS, 55, 389
\bibitem[\protect\citeauthoryear{Nardetto et al}{2004}]{nardetto2004}
Nardetto N., Fokin A., Mourard D., Mathias Ph., Kervalla P., Bersier
D., 2004, A \& A, 428, 131
\bibitem[\protect\citeauthoryear{Petterson}{2002}]{petterson2002}
Petterson O., 2002, PhD thesis, University of Canterbury
\bibitem[\protect\citeauthoryear{Pollard et al.}{1996}]{pollard1996} Pollard
K.R., Cottrell P.L., Kilmartin P., Gilmore A., 1996a, MNRAS, 279, 949
\bibitem[\protect\citeauthoryear{Sabbey et al.}{1995}]{sabbey1995}
Sabbey C.N., Sasselov D.D., Fieldus M.S., Lester J.B., Venn K.A.,
Butler R.P., 1995, ApJ, 446, 250
\bibitem[\protect\citeauthoryear{Schmidt et al.}{2003a}]{schmidt2003}
Schmidt E.G., Lee K.M., Johnston D., Newman P.R., Snedden S.A., 2003,
AJ, 126, 906
\bibitem[\protect\citeauthoryear{Schmidt et al.}{2003b}]{schmidt2003ii}
Schmidt E.G., Langan S., Lee K.M., Johnston D., Newman P.R., Snedden
S.A.. 2003, AJ, 126, 2495
\bibitem[\protect\citeauthoryear{Vinko et al.}{1998}]{vinko1998}
Vink\'{o} J., Evans N., Kiss L., Szabados L., 1998, MNRAS, 296, 824
\bibitem[\protect\citeauthoryear{Wallerstein et
al.}{1992}]{wallerstein1992} Wallerstein G., Jacobsen T.S., Cottrell
P.L., Clark M., Albrow M., 1992, MNRAS, 259, 474

\end{thebibliography}
\end{document}